\documentclass[12pt]{article}
\begin{document}

\title{{\Large \textbf{Self-similar Bianchi type VIII and IX models}}}
\author{Pantelis S. Apostolopoulos and Michael Tsamparlis \\
{\small \textit{{Department of Physics, Section of
Astrophysics-Astronomy-Mechanics},}}\\
{\small \textit{University of Athens, Panepistemiopolis, Athens 157 83,
Greece}}}
\maketitle

\begin{abstract}
It is shown that in transitively self-similar spatially homogeneous tilted
perfect fluid models the symmetry vector is not normal to the surfaces of
spatial homogeneity. A direct consequence of this result is that there are
no self-similar Bianchi VIII and IX tilted perfect fluid models. Furthermore
the most general Bianchi VIII and IX spacetime which admit a four
dimensional group of homotheties is given.

\textbf{PACS: 4.20.-q, 4.20.Ha}
\end{abstract}

\section{Introduction}

Despite the fact that in Spatially Homogeneous (SH) models the field
equations are reduced to a system of ordinary differential equations, not
many exact solutions are known, especially in the case of tilted perfect
fluid models. This has initiated the study of these models using the methods
of the theory of dynamical systems, where one examines their behaviour from
a qualitative point of view and, in particular, at early, late and
intermediate periods of their history. These studies have revealed that 
\emph{transitively self-similar} SH models act as early (i.e. near to the
initial singularity) and late time asymptotic states for more general
spatially homogeneous models \cite{Wainwright-Ellis}. However it has been
pointed out that there is an open set of SH models which are not
asymptotically self-similar i.e. \emph{may not} admit proper HVF (although,
in some cases, their asymptotic states can be successively approximated by
an infinite sequence of self-similar models). For these and other reasons it
is of primary interest to determine all SH perfect fluid models which are or
not \emph{transitively self-similar}. 

We recall that a self-similar model admits a Homothetic Vector Field (HVF) $%
\mathbf{H}$ and is defined by the requirement: 
\begin{equation}
\mathcal{L}_{\mathbf{H}}g_{ab}=2\psi g_{ab}  \label{sx1}
\end{equation}
where $\psi $=const. is the homothetic factor.

Self-similar SH vacuum and non-tilted perfect fluid models have been
determined by Hsu and Wainwright \cite{Hsu-Wainwright} who proved that
Bianchi type VIII and IX models are not self-similar \cite
{Wainwright-Ellis,Hsu-Wainwright}. Concerning the tilted perfect fluid
models, Bradley \cite{Bradley} has stated that there do not exist tilted
dust self-similar models whereas Hewitt \cite{Hewitt} has found the general
self-similar Bianchi type II solution and Rosquist and Jantzen have
determined a rotational tilted Bianchi VI$_0$ self-similar model \cite
{Rosquist1,Rosquist-Jantzen1}.

The Bianchi II solution found by Hewitt is \emph{unique} amongst the tilted
Bianchi models which prossess a $G_{2}$ (abelian) subgroup acting
orthogonally transitively \cite{Hewitt}. Nevertheless, because Bianchi type
VIII and IX do not have this property, the question if tilted perfect fluid
Bianchi type VIII and IX models admit a proper HVF is still open.

In this paper we prove that the answer to this question is negative, that
is, tilted perfect fluid Bianchi type VIII and IX models do not admit a
proper HVF. 

It is important to state clearly our assumptions, which are as follows:

a) The spacetime manifold admits a $G_{3}$ group of isometries acting simply
transitively on 3-dimensional spacelike surfaces $\mathcal{O}$. It is well
known \cite{Ellis-MacCallum} that the unit normal $u^{a}$ ($u^{a}u_{a}=-1$)
to the surfaces of homogeneity $\mathcal{O}$ is geodesic and rotation free
i.e. $u_{[a;b]}=\omega _{ab}=0$.

b) The matter content of the SH model is tilted perfect fluid moving with
4-velocity $\tilde{u}^{a}$ ($\tilde{u}^{a}\tilde{u}_{a}=-1$) which is not
orthogonal to the surfaces of homogeneity.

c) The SH model is \emph{transitively self-similar}, that is, it admits a
proper HVF $\mathbf{H}$ which, together with the Killing Vectors, generate a
simply transitive homothety group of transformations of the spacetime
manifold.

\section{Self-similar Bianchi models}

To establish the relation between the dynamic quantities defined by the
observers $u^{a}$,$\tilde{u}^{a}$ we consider the 1+3 decomposition of the
energy momentum tensor induced by each of them \cite{Ellis1}. For the field $%
u^{a}$ one has:

\begin{equation}
T_{ab}=\mu u_{a}u_{b}+ph_{ab}+2q_{(a}u_{b)}+\pi _{ab}  \label{sx3}
\end{equation}
where the dynamical quantities $\mu $, $p$, $q_{a}$ and $\pi _{ab}$ are
defined as follows:

\begin{equation}
\mu =T_{ab}u^{a}u^{b},\mbox{ }p=\frac{1}{3}T_{ab}h^{ab},\mbox{ }%
q_{a}=-h_{a}^{c}T_{cd}u^{d},\pi _{ab}=h_{a}^{c}h_{b}^{d}T_{cd}-\frac{1}{3}%
(h^{cd}T_{cd})h_{ab}  \label{sx4}
\end{equation}
and $h_{ab}=g_{ab}+u_{a}u_{b}$ being the projection tensor associated with $%
u^{a}$.

For the tilted 4-velocity $\tilde{u}_{a}$ we have assumed that:

\begin{equation}
T_{ab}=\tilde{\mu}\tilde{u}_{a}\tilde{u}_{b}+\tilde{p}\tilde{h}_{ab}
\label{sx5}
\end{equation}
where $\tilde{\mu},\tilde{p}$ are the energy density and isotropic pressure
respectively, measured by the tilted observers $\tilde{u}^{a}$ and $\tilde{h}%
_{ab}=g_{ab}+\tilde{u}_{a}\tilde{u}_{b}$ is the projection tensor of $\tilde{%
u}^{a}$.

Comparing (\ref{sx4}) and (\ref{sx5}) one obtains the following relations
among the corresponding quantities \cite{van Elst-Uggla}:

\[
\mu =\tilde{\mu}+\Gamma ^{2}\upsilon ^{2}(\tilde{\mu}+\tilde{p}) 
\]
\[
p=\tilde{p}+\frac{1}{3}\Gamma ^{2}\upsilon ^{2}(\tilde{\mu}+\tilde{p}) 
\]

\begin{equation}
q_{a}=\Gamma ^{2}(\tilde{\mu}+\tilde{p})\upsilon _{a}  \label{sx8}
\end{equation}
\begin{equation}
\pi _{ab}=\Gamma ^{2}(\tilde{\mu}+\tilde{p})\left( \upsilon _{a}\upsilon
_{b}-\frac{1}{3}\upsilon ^{2}h_{ab}\right)  \label{sx9}
\end{equation}
where: 
\begin{equation}
\tilde{u}_{a}=\Gamma \left( u_{a}+\upsilon _{a}\right)  \label{sx10}
\end{equation}
\[
\Gamma =\left( 1-\upsilon ^{2}\right) ^{-\frac{1}{2}} 
\]
and $\upsilon ^{2}=\upsilon ^{a}\upsilon _{a}$, $u^{a}\upsilon _{a}=0$.

We come now to the kinematical and the dynamical implications of the
existence of the HVF. Concerning the kinematic part one has the following,
easily established, result \cite{Oliver-Davis1,
Maartens-Mason-Tsamparlis1,Coley-CKV}:

\textbf{Proposition 1}

\textit{A spacetime admits a timelike HVF }$H^a=Hn^a$\textit{\ parallel to
the unit timelike vector field }$n^a$\textit{\ (}$n^an_a=-1$\textit{) iff: } 
\[
\sigma _{ab}=0
\]
\[
\dot{n}_a=\left( \ln H\right) _{;a}+\frac \theta 3n_a.
\]
\textit{Moreover the homothetic factor }$\psi $\textit{\ satisfies: } 
\[
\psi =\frac{H\theta }3
\]
\textit{where }$\sigma _{ab}=\left( h_a^ch_b^d-\frac 13h_{ab}h^{cd}\right)
n_{(c;d)}$\textit{\ is the shear tensor, }$\dot{n}_a=n_{a;b}n^b$\textit{\ is
the acceleration, }$\theta =h^{cd}n_{c;d}$\textit{\ is the rate of expansion
and }$h_{ab}$\textit{\ is the associated projection tensor of the timelike
congruence }$n^a$\textit{\ \cite{Ellis1}. }

From Proposition 1 one has the useful result that if a SH model admits a
proper HVF parallel to the fluid velocity $n^{a}=u^{a}$ then $\sigma _{ab}=0$%
. Because the timelike congruence $u^{a}$ is also geodesic and irrotational,
the ($0\alpha $)-constraint equation and the $H_{ab}$-equation \cite
{Ellis1,Coley-McManus1} imply:

\[
\frac{2}{3}h_{b}^{a}\theta ^{;b}=q^{a}\mbox{ and }H_{ab}=0 
\]
where $H_{ab}$ is the magnetic part of the Weyl tensor.

Due to the spatial homogeneity of the Bianchi models $h_b^a\theta
^{;b}=0=q^a $ therefore, from equation (\ref{sx10}), it follows that $u^a=%
\tilde{u}^a$ and the fluid is necessarily non-tilted. Because the only such
models are the FRW cosmological models \cite{Wainwright-Ellis}, we conclude
that \emph{the only perfect fluid Bianchi spacetimes which admit a proper
HVF parallel to }$u^a$\emph{\ are the corresponding FRW models} \cite
{Hsu-Wainwright}. It is interesting to note that the above result extends to
the more general case of a proper CKV (i.e. $\psi _{;a}\neq 0$). Hence \emph{%
perfect fluid Bianchi spacetimes do not admit proper CKVs or HVF parallel to
the unit normal }$u^a$,\emph{\ except their FRW analogues }which is in
agreement with the result of Coley and Tupper \cite{Coley-Tupper Inheriting
CKV}.

Using the above arguments we prove the main result of the paper: \pagebreak

\textbf{Proposition 2}

\textit{There are no Bianchi type VIII and IX transitively self-similar
tilted perfect fluid models. }

\textbf{Proof}

Suppose that Bianchi type VIII, IX models are transitively self-similar i.e.
they admit a homothety group $H_{4}$ with 4-dimensional orbits and
generators $\{\mathbf{H},\mathbf{X}_{\alpha }\}$ where $\{\mathbf{X}_{\alpha
}\}$ (greek indices take the values $1,2,3$) are the generators of the $%
G_{3} $ group of isometries acting simply transitively on the spacelike
hypersurfaces $\mathcal{O}$ and $\mathbf{H}$ is the HVF. The Jacobi
identities applied to the vector fields $\{\mathbf{H},\mathbf{X}_{\alpha }\}$
imply that $\mathbf{H}$ is invariant under $G_{3}$ \cite{MacCallum1,koutras}%
: 
\begin{equation}
\lbrack \mathbf{X}_{\alpha },\mathbf{H}]=0.  \label{sx11}
\end{equation}
We decompose $\mathbf{H}$ along and perpendicular to $\mathbf{u}$ as
follows: 
\begin{equation}
H^{a}=Hu^{a}+Y^{a}  \label{sx12}
\end{equation}
where $H=-H^{a}u_{a}$ and $u^{a}Y_{a}=0$.

The commutator (\ref{sx11}) gives $\mathbf{X}_{\alpha }(H)\mathbf{u}=[%
\mathbf{Y},\mathbf{X}_{\alpha }]$. Because $\mathbf{Y}$ lies in the
hypersurfaces of spatial homogeneity, it follows that $\mathbf{X}_{\alpha
}(H)=0$ and $[\mathbf{Y},\mathbf{X}_{\alpha }]=0$, that is, both $\mathbf{Y}$
and $H$ are invariant under the action of $G_{3}$. Furthermore, since $%
\mathbf{H}$ is a HVF of the spacetime manifold and $\mathbf{u}$ is geodesic
and irrotational, one has $[\mathbf{H,u]}=-\mathbf{u}(H)\mathbf{u+[Y},%
\mathbf{u]=}-\psi \mathbf{u}$ where $\mathbf{u}(H)=\psi $ \cite
{Maartens-Mason-Tsamparlis1} from which it follows $\mathbf{[Y},\mathbf{u]=}%
0 $ i.e. $\mathbf{Y}$ is invariant under $\mathbf{u}$.

Therefore from (\ref{sx12}) we may write $\mathbf{H}=H\mathbf{u}+A^\alpha 
\bf{\eta }_\alpha $ where $\bf{\eta }_\alpha $ is the
group-invariant basis \cite{MacCallum2} and $A^\alpha $ are \emph{constants}%
. Because $[\mathbf{H},\bf{\eta }_\alpha ]=0$ it follows that $A^\alpha
C_{\alpha \beta }^\gamma =0 $ where $C_{\alpha \beta }^\gamma $ are the
structure constants of the Bianchi type VIII and IX group of isometries.
Hence $A^\alpha =0$ which, in turn, implies that $\mathbf{H}=H\mathbf{u}$
and the HVF $\mathbf{H}$ is parallel to $\mathbf{u}$. By virtue of the
previous considerations we conclude that self-similar Bianchi VIII, IX
models do not admit a tilted (or non-tilted) perfect fluid interpretation,
unless the spacetime becomes a FRW spacetime in which case the Bianchi IX is
the only permissible type.

A side result of Proposition 2 is that the \emph{general solution} of the
homothetic equations in Bianchi type VIII and IX spacetimes is: 
\begin{equation}
ds^2=-dt^2+(\psi t)^2g_{\alpha \beta }\mathbf{\omega }^\alpha \mathbf{\omega 
}^\beta  \label{sx19}
\end{equation}
where $g_{\alpha \beta }$ are constants of integration and $\mathbf{\omega }%
^\alpha $ are the Bianchi VIII or IX invariant 1-forms \cite{MacCallum2}.

In this case the HVF $H^a=\psi t\delta _t^a$ and the fluid interpretation of
self-similar Bianchi VIII and IX\ spacetimes necessarily involve anisotropic
stress i.e. $\pi _{ab}\neq 0$, whether the fluid is tilted or non-tilted.

We note that the structure of a four dimensional Lie Algebra \cite
{MacCallum1} shows that amongs the Bianchi models sharing the property of the 
existence of a proper HVF, the ''singular'' behaviour of $H_4$ (equation (\ref{sx11}))
appears only in Bianchi types VIII and IX. Therefore we expect that this
result does not extend to the rest of Bianchi models and the method
developed in this paper can be easily applied leading towards to the
determination of the homothety group $H_4$ and the structure of tilted
perfect fluid Bianchi spacetimes. This will be the subject of a subsequent
work.


\begin{thebibliography}{99}
\bibitem{Wainwright-Ellis}  Wainwright J and Ellis G F R (eds), \emph{%
Dynamical Systems in Cosmology }(Cambridge University Press, Cambridge 1997).

\bibitem{Hsu-Wainwright}  Hsu L and Wainwright J, \emph{Self-Similar
spatially homogeneous cosmologies: orthogonal perfect fluid and vacuum
solutions}, 1986 Class. Quantum Grav. \textbf{3} 1105-1124.

\bibitem{Bradley}  Bradley M, \emph{Dust EPL cosmologies}, 1988 Class.
Quantum Grav. \textbf{5} L15-L19.

\bibitem{Hewitt}  Hewitt C G, \emph{An exact tilted Bianchi II cosmology},
1991 Class. Quantum Grav. \textbf{8} L109-L114.

\bibitem{Rosquist1}  Rosquist K, \emph{Exact rotating and expanding
radiation-filled universe}, 1983 Phys. Lett. \textbf{97}A 145-146.

\bibitem{Rosquist-Jantzen1}  Rosquist K and Jantzen R T, \emph{Exact power
law solutions of the Einstein equations}, 1985 Phys. Lett. \textbf{107}A
29-32.

\bibitem{Ellis-MacCallum}  Ellis G F R and MacCallum M A H, \emph{A class of
homogeneous cosmological models}, 1969 Commun. Math. Phys. \textbf{12}
108-141.

\bibitem{Ellis1}  Ellis G F R 1973 \emph{Carg\`{e}se Lectures in Physics,
Vol. 6, }Ed. E Schatzman (Gordon and Breach, New York).

\bibitem{van Elst-Uggla}  van Elst H and Uggla C, \emph{General relativistic
1+3 orthonormal frame approach revisited}, 1997 Class. Quantum Grav. \textbf{%
14} 2673-2695.

\bibitem{Oliver-Davis1}  Oliver D R and Davis W R, \emph{On certain timelike
symmetry properties and the evolution of matter fluid spacetimes that admit
them}, 1977 Gen. Rel. Grav. \textbf{8} 905-908.

\bibitem{Maartens-Mason-Tsamparlis1}  Maartens R, Mason D P and Tsamparlis
M, \emph{Kinematic and dynamic properties of conformal Killing vectors in
anisotropic fluids}, 1986 J. Math. Phys. \textbf{27} 2987-2994.

\bibitem{Coley-CKV}  Coley A A, \emph{Fluid spacetimes admitting a conformal
Killing vector parallel to the velocity vector}, 1991 Class. Quantum Grav. 
\textbf{8} 955-968.

\bibitem{Coley-McManus1}  Coley A A and McManus D J, \emph{On space-times
admitting shear-free, irrotational, geodesic timelike congruences}, 1994
Class. Quantum Grav. \textbf{11} 1261-1282.

\bibitem{Coley-Tupper Inheriting CKV}  Coley A A and Tupper B O J, \emph{%
Spacetimes admitting inheriting conformal Killing vector fields}, 1990 Class.
Quantum Grav. \textbf{7} 1961-1981.

\bibitem{MacCallum1}  MacCallum M A H, \emph{On the classification of the
real four-dimensional Lie Algebras}, 1979 Queen Mary and Westfield College
preprint.

\bibitem{koutras}  Koutras A 1992 \emph{Mathematical properties of
homothetic space-times}, Ph.D. thesis, Queen Mary and Westfield College.

\bibitem{MacCallum2}  MacCallum M A H 1979, \emph{Anisotropic and
inhomogeneous relativistic cosmologies}\textit{, }General Relativity, Eds S.
W. Hawking and W. Israel \textit{(}Cambridge University Press, Cambridge).
\end{thebibliography}
\end{document}